\documentclass[showpacs,amsmath,amssymb,twocolumn,prl,superscriptaddress]{revtex4-1}
\usepackage{amssymb}
\usepackage[dvips]{graphicx}
\usepackage{enumerate}
\usepackage{epsfig}
\usepackage{subfigure}
\usepackage{xcolor}
\usepackage[T1]{fontenc}
\usepackage[expert]{mathdesign}
\usepackage{fullpage}
\usepackage{amsthm,amsfonts,amssymb,amscd,mathrsfs,eufrak,xspace,framed}
\usepackage{amsmath}
\usepackage{color}
\usepackage{setspace}
\usepackage{url}
\usepackage{wrapfig}
\usepackage{enumitem}
\usepackage{comment}

\begin{document}

\title{High-Purity Pulsed Squeezing Generation with Integrated Photonics}

\author{Chaohan Cui}
\affiliation{James C. Wyant College of Optical Sciences, The University of Arizona, Tucson, Arizona 85721, USA}
\author{Christos N. Gagatsos}
\affiliation{James C. Wyant College of Optical Sciences, The University of Arizona, Tucson, Arizona 85721, USA}
\author{Saikat Guha}
\affiliation{James C. Wyant College of Optical Sciences, The University of Arizona, Tucson, Arizona 85721, USA}
\author{Linran Fan}
\email{lfan@optics.arizona.edu}
\affiliation{James C. Wyant College of Optical Sciences, The University of Arizona, Tucson, Arizona 85721, USA}

\begin{abstract}
Squeezed light has evolved into a powerful tool for quantum technology, ranging from quantum enhanced sensing and quantum state engineering based on partial post-selection techniques. The pulsed generation of squeezed light is of particular interest, as it can provide accurate time stamp and physically defined temporal mode, which are highly preferred in complex communication networks and large-scale information processing. However, the multimode feature of pulsed squeezing in conventional single-pass configuration limits the purity of the output state, 
negatively impacting its application in quantum technology. In this Letter, we propose a new approach to generate pulsed squeezing with high temporal purity. Pulsed squeezing based on parametric down-conversion in photonic cavities is analyzed. We show that the effective mode number of the output squeezed light approaches unity. Such a high-purity squeezed light can be realized with broad parameters and low pump power, providing a robust approach to generate large-scale quantum resource.
\end{abstract}

\maketitle

\section{Introduction}
Non-Gaussian states are indispensable resources required by quantum information processing to demonstrate quantum advantage \cite{weedbrook2012gaussian}. Partial detection of squeezed light is one of the most important optical approach to generate non-Gaussian states \cite{opatrny2000improvement,ourjoumtsev2007increasing,wakui2007photon,namekata2010non,walschaers2017entanglement}. Optical cat and kitten states have been generated based on photon-subtraction from single-mode squeezed vacuum \cite{ralph2003quantum,ourjoumtsev2006generating,neergaard2006generation,asavanant2017generation}. In principle, arbitrary non-Gaussian states, including Gottesman-Kitaev-Preskill (GKP) state for cluster-modal quantum computing \cite{GKP,menicucci2006universal,menicucci2014fault,asavanant2019generation,baragiola2019all,pfister2019continuous,cui2020high}, can be generated based on Gaussian-Boson-Sampling (GBS) configuration and photon-number-resolving (PNR) detection \cite{PNRstateengineering1,PNRstateengineering2}. One critical requirement to implement partial detection of squeezed light is that all photons must be in the same spectral-temporal mode. Otherwise, unconditioned Gaussian modes will be mixed with the target non-Gaussian mode, thus decrease the purity of the output state. Common techniques utilized for single photons, such as spectral filtering and post-selection within small time window, cannot be applied for squeezed light due to the excess loss. 

The standard configuration to implement pulsed squeezing is single-pass parametric down-conversion \cite{slusher1987pulsed,kim1994quadrature,eto2007observation,eckstein2011highly,andersen201630}. This process intrinsically involves multiple modes in both space and time, which all have significant squeezing and energy \cite{wasilewski2006pulsed}. Synchronously-pumped parametric down-conversion in free-space cavities has also been proposed for pulsed squeezing \cite{patera2010quantum, jiang2012time}. However, the generated squeezing still contains significant multi-mode contribution. It also requires that the pump repetition rate matches cavity free-spectra-range (FSR), which is challenging for integrated cavities with large FSR. While complex shaping of local oscillator can be utilized to improve the measured squeezing level \cite{eto2008observation}, it does not work well on non-Gaussian state generation through partial detection, which requires the separation of different modes. Therefore, a new approach to generate pulsed squeezing with high purity is highly desired to future improve the capability of photonic quantum information processing.


In this Letter, we propose a novel approach to generate pulsed squeezing with high purity. Parametric down-conversion in integrated photonic cavities with second-order nonlinearity is pumped with a pulsed source.  Bloch-Messiah approach is used to decompose the input-output relation into independent squeezing modes \cite{bloch1962canonical,braunstein2005squeezing}. We demonstrate that the effective mode number at the output can approach unity, showing there is only one dominate spectral-temporal mode. Unlike single-pass pulsed squeezing \cite{wasilewski2006pulsed}, this approach does not require delicate matching between pump bandwidth and amplitude, making it robust and promising for the generation of complex quantum states.

\section{Theoretical model}
The proposed configuration is shown in Fig. 1, which consists of a photonic ring cavity evanescently couped to a bus waveguide. Phase matching condition is satisfied for degenerate parametric down-conversion between the pump mode at frequency $2\omega_0$ and signal mode at frequency $\omega_0$. The input pump pulse with center frequency $\omega_{p0}=2\omega_0$ is launched into the cavity through the bus waveguide.

\begin{figure}[h]
\centering
\includegraphics[width=2 in]{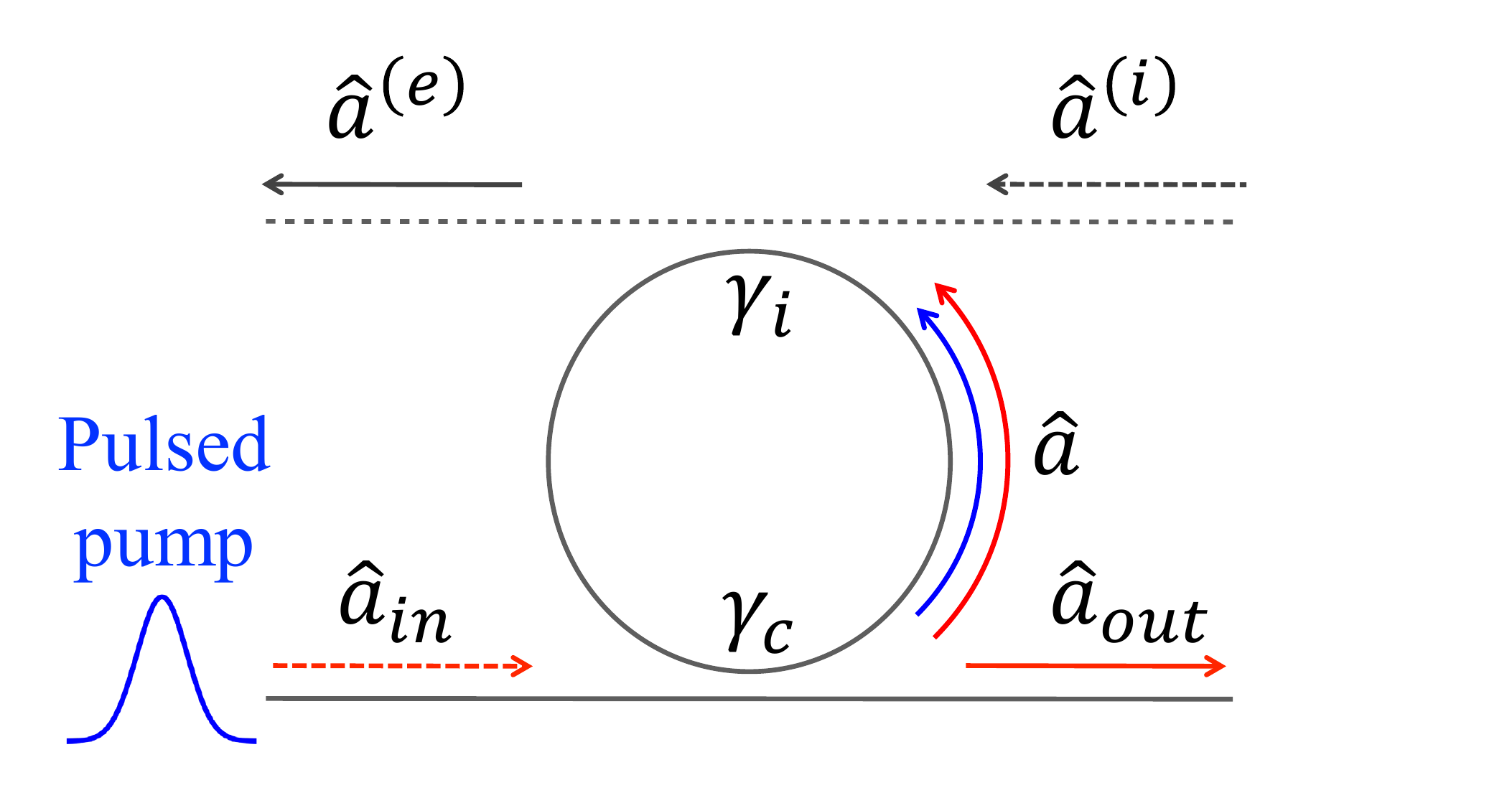}
\caption{\label{fig:ThDiagram1} Proposed configuration for high-purity pulsed squeezing generation: parametric down-conversion in photonic ring cavities with pulsed pump.}
\end{figure}

The dynamics of the intra-cavity signal mode  $\hat{a}$ can be described by the equation of motion\cite{walls2007quantum,aspelmeyer2014cavity},
\begin{equation}
\label{EoM}
    \frac{d\hat{a}}{dt}=\frac{i}{\hbar}[\hat{H},\hat{a}]-\frac{\gamma}{2}\hat{a}+\sqrt{\gamma_i}\hat{a}^{(i)}+\sqrt{\gamma_c}\hat{a}_{in}
\end{equation}
with intrinsic loss rate $\gamma_i$, bus waveguide coupling rate $\gamma_c$, total cavity decay $\gamma=\gamma_i+\gamma_c$, and noise operators due to intrinsic loss $\hat{a}^{(i)}$ and bus waveguide $\hat{a}_{in}$. The Hamiltonian $\hat{H}$ can be written as
\begin{equation}
\begin{aligned}
    \hat{H}=&\hbar\omega_0 \int d\omega \hat{a}^\dag(\omega) \hat{a}(\omega) \\
    &+\frac{i\hbar\kappa}{2}\iint d\omega d\omega'\hat{a}^\dag(\omega)\hat{a}^\dag(\omega')\varepsilon(\omega+\omega') +h.c.
\end{aligned}
\end{equation}
with $\kappa$ the single photon coupling rate for parametric down-conversion. The intra-cavity pump field $\varepsilon(\omega)$ is written as
\begin{equation}
    \varepsilon(\omega)= E_p(\omega)\frac{\sqrt{\gamma_{pc}}}{-i(\omega-2\omega_{0})+\gamma_p/2}
\end{equation}
with the bus waveguide coupling rate $\gamma_{pc}$ and total decay rate $\gamma_p$ for the pump mode, and $E_p(\omega)$ the spectrum amplitude of the input pulse. Utilizing Fourier transformation, Eq.~\eqref{EoM} can be further expressed in frequency domain
\begin{equation}
\label{eq:ME}
\begin{aligned}
    0=&\int d\omega' [i(\omega-\omega_0)-\frac{\gamma}{2}]\delta(\omega-\omega')\hat{a}(\omega')\\
    &+\int d\omega' \kappa\varepsilon(\omega+\omega')\hat{a}^\dag(\omega')\\
    &+\sqrt{\gamma_i}\hat{a}^{(i)}(\omega)+\sqrt{\gamma_c}\hat{a}_{in}(\omega).
\end{aligned}
\end{equation}
By including the complex-conjugation, we can rewrite Eq.~\eqref{eq:ME} into the following equivalent matrix form
\begin{equation}
\begin{aligned}
    0=&\begin{pmatrix}
    D & E\\ 
    E^\dag & D^\dag
    \end{pmatrix}
    \begin{pmatrix}
    \hat{a}(\omega)\\ 
    \hat{a}^\dag(\omega)
    \end{pmatrix}\\
    &+\sqrt{\gamma_i}\begin{pmatrix}
    \hat{a}^{(i)}(\omega)\\ 
    \hat{a}^{(i)\dag}(\omega)
    \end{pmatrix}+\sqrt{\gamma_c}\begin{pmatrix}
    \hat{a}_{in}(\omega)\\ 
    \hat{a}_{in}^\dag(\omega)
    \end{pmatrix}
\end{aligned}
\end{equation}
with the diagonal matrix $D(\omega,\omega')=[i(\omega-\omega_0)-\gamma/2]\cdot\delta(\omega-\omega')$ shows the effect of frequency detuning and energy decay, and the matrix $E(\omega,\omega')=\kappa\varepsilon(\omega+\omega')$ shows the nonlinear interaction. Then the output field can be derived based on the input-output theory

\begin{equation}
\begin{aligned}
\label{IO}
\begin{pmatrix}
\hat{a}_{out}(\omega) \\ 
\hat{a}_{out}^\dag(\omega)
\end{pmatrix}=&\left[\begin{pmatrix}
I & \\ 
 & I
\end{pmatrix}+\gamma_c
\begin{pmatrix}
D & E\\ 
E^\dag & D^\dag
\end{pmatrix}^{-1}\right]
\begin{pmatrix}
\hat{a}_{in}(\omega) \\ 
\hat{a}_{in}^\dag(\omega)
\end{pmatrix}\\
&+\sqrt{\gamma_c\gamma_i}
\begin{pmatrix}
D & E\\ 
E^\dag & D^\dag
\end{pmatrix}^{-1}\begin{pmatrix}
\hat{a}^{(i)}(\omega) \\ 
\hat{a}^{(i)\dag}(\omega)
\end{pmatrix}.
\end{aligned}
\end{equation}

Here we assume the intrinsic loss of the photonic cavity is Markovian \cite{aspelmeyer2014cavity}, and can be modeled as a virtual waveguide with the input mode $\hat{a}^{(i)}$ and output mode $\hat{a}^{(e)}$ (dashed line in Fig.~\ref{fig:ThDiagram1}). Then Eq.~\eqref{IO} is converted to the symplectic form
\begin{widetext}
\begin{equation}
\begin{aligned}
\label{Decompose}
\begin{pmatrix}
\hat{a}_{out}(\omega) \\ 
\hat{a}_{out}^\dag(\omega) \\
\hat{a}^{(e)}(\omega) \\ 
\hat{a}^{(e)\dag}(\omega)
\end{pmatrix}=&\begin{pmatrix}\left[\begin{pmatrix}
I & \\ 
 & I
\end{pmatrix}+\gamma_c
\begin{pmatrix}
D & E\\ 
E^\dag & D^\dag
\end{pmatrix}^{-1}\right] & \sqrt{\gamma_c\gamma_i}\begin{pmatrix}
D & E\\ 
E^\dag & D^\dag
\end{pmatrix}^{-1}\\
\sqrt{\gamma_c\gamma_i}\begin{pmatrix}
D & E\\ 
E^\dag & D^\dag
\end{pmatrix}^{-1} & \left[\begin{pmatrix}
I & \\ 
 & I
\end{pmatrix}+\gamma_i
\begin{pmatrix}
D & E\\ 
E^\dag & D^\dag
\end{pmatrix}^{-1}\right] \end{pmatrix}
\begin{pmatrix}
\hat{a}_{in}(\omega)\\ 
\hat{a}_{in}^\dag(\omega)\\
\hat{a}^{(i)}(\omega)\\ 
\hat{a}^{(i)\dag}(\omega)
\end{pmatrix}\\
=&\begin{pmatrix}
\sqrt{\frac{\gamma_c}{\gamma}} & -\sqrt{\frac{\gamma_i}{\gamma}} \\ 
\sqrt{\frac{\gamma_i}{\gamma}} & \sqrt{\frac{\gamma_c}{\gamma}}
\end{pmatrix}\begin{pmatrix}\left[\begin{pmatrix}
I & \\ 
 & I
\end{pmatrix}+\gamma
\begin{pmatrix}
D & E\\ 
E^\dag & D^\dag
\end{pmatrix}^{-1}\right] & 0\\
0 & \begin{pmatrix}
I & \\ 
 & I
\end{pmatrix} \end{pmatrix}
\begin{pmatrix}
\sqrt{\frac{\gamma_c}{\gamma}} & \sqrt{\frac{\gamma_i}{\gamma}} \\ 
-\sqrt{\frac{\gamma_i}{\gamma}} & \sqrt{\frac{\gamma_c}{\gamma}}
\end{pmatrix}\begin{pmatrix}
\hat{a}_{in}(\omega)\\ 
\hat{a}_{in}^\dag(\omega)\\
\hat{a}^{(i)}(\omega)\\ 
\hat{a}^{(i)\dag}(\omega)
\end{pmatrix}.
\end{aligned}
\end{equation}
Based on Bloch-Messiah decomposition \cite{bloch1962canonical,braunstein2005squeezing}, the core matrix $\left[\begin{pmatrix}
I & \\ 
 & I
\end{pmatrix}+\gamma
\begin{pmatrix}
D & E\\ 
E^\dag & D^\dag
\end{pmatrix}^{-1}\right]$ 
can be decomposed into a series of independent Bogoliubov transformations with unitary matrix $P$ and $Q$.
\begin{equation}\left[\begin{pmatrix}
I & \\ 
 & I
\end{pmatrix}+\gamma
\begin{pmatrix}
D & E\\ 
E^\dag & D^\dag
\end{pmatrix}^{-1}\right]=
\begin{pmatrix}
P & \\
& P^*
\end{pmatrix}
\begin{pmatrix}
 \cosh\xi_1&  &  &e^{i\theta_1}\sinh\xi_1  &  & \\ 
 &  ... &  &  &...  & \\ 
 &  &   \cosh\xi_n&  &  &e^{i\theta_n}\sinh\xi_n \\ 
e^{-i\theta_1}\sinh\xi_1 &  &  &   \cosh\xi_1&  & \\ 
 & ... &  &  &  ...& \\ 
 &  &e^{-i\theta_n}\sinh\xi_n  &  &  &  \cosh\xi_n
\end{pmatrix}
\begin{pmatrix}
Q^\dag & \\
& Q^T
\end{pmatrix}.
\end{equation}
\end{widetext}

\begin{figure}[htb]
\centering
\includegraphics[width=7.8cm]{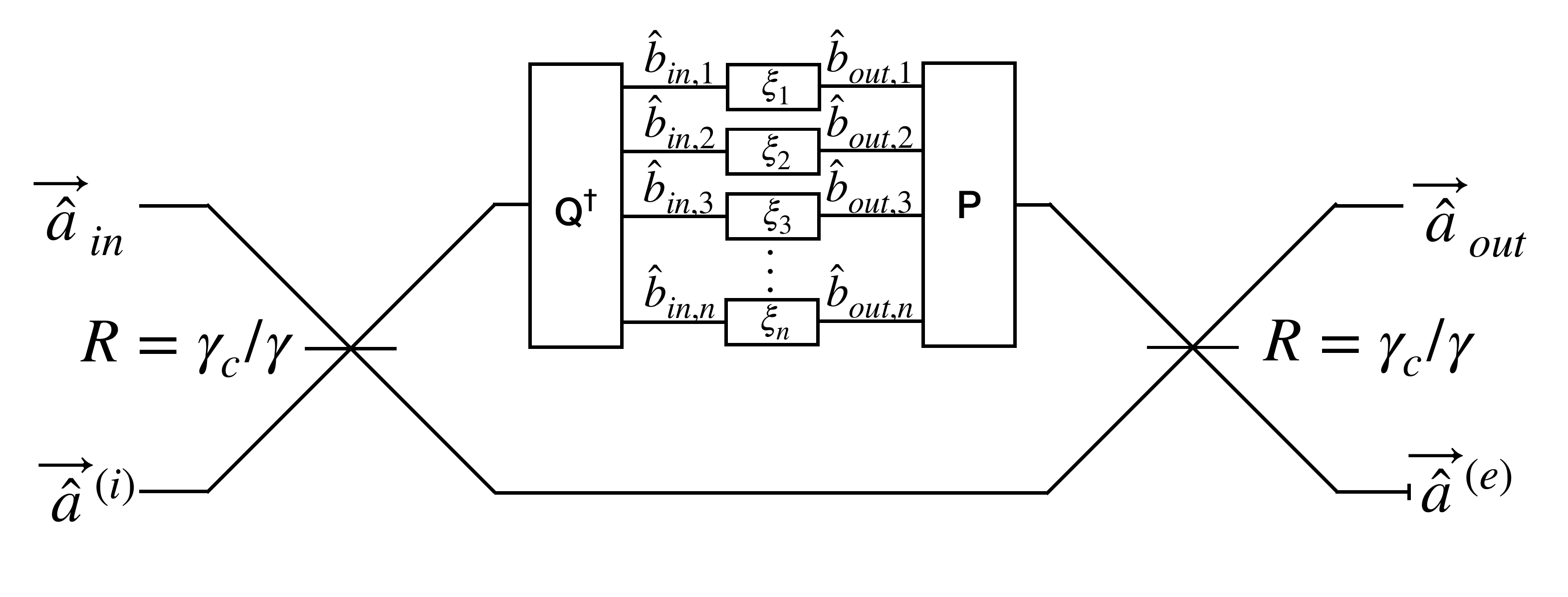}
\caption{\label{fig:ThDiagram2} Equivalent photonic circuit of parametric down-conversion in a photonic cavity with pulsed pump.}
\end{figure}

Therefore, the overall input-output relation is modeled as a multi-mode optical parametric amplifier sandwiched by two beamsplitters with reflection $R=\gamma_c/\gamma$ (Fig.~\ref{fig:ThDiagram2}). The spectral-temporal shape of the characteristic modes (${b}_{in,k}$) is determined by the unitary transformation $Q$.
\begin{equation}
\hat{b}_{in} = Q^\dag\left( \sqrt{\frac{\gamma_c}{\gamma}}\hat{a}_{in}(\omega) + \sqrt{\frac{\gamma_i}{\gamma}}\hat{a}^{(i)}(\omega)\right).
\end{equation}
Each characteristic mode undergoes independent squeezing
\begin{equation}
\hat{b}_{out,k} = \cosh\xi_k \hat{b}_{in,k}+e^{i\theta_k}\sinh\xi_k\hat{b}_{in,k}^\dag
\end{equation}
where $\xi_k$ and $\theta_k$ are the squeezing amplitude and phase of the $k$th mode. After mixing with vacuum at the second beamsplitter, the variance of the squeezed quadrature of the $k$th mode is
\begin{equation}
\label{Var}
\langle(\Delta X_k(\frac{\theta_k-\pi}{2}))^2\rangle = \frac{1}{2}(\frac{\gamma_i}{\gamma}+\frac{\gamma_c}{\gamma}e^{-2\xi_k}).
\end{equation}
This result is identical to the well-known CW squeezing \cite{walls2007quantum}, where the maximum squeezing is limited by the intrinsic loss of the cavity. Over-coupled cavity ($\gamma_c>>\gamma_i$) is required to realize high squeezing. The effective mode number $K$ can be directly calculated from the squeezing amplitude $\xi_k$.
\begin{equation}
\label{Mode_number}
\begin{aligned}
K=\frac{(\sum_k \sinh^2\xi_k)^2}{\sum_k \sinh^4\xi_k}
\end{aligned}
\end{equation}
The effective mode number equals to unity when there is only one non-zero squeezing amplitude. 

Our derivation assumes that the pump is below threshold so that all the output modes are still squeezed vacuum. The pump threshold can be determined when the gain for any intra-cavity mode is equal to the amplitude loss rate $\gamma/2$. By rewriting Eq.~\eqref{EoM} into rotation frame $\hat{a} \rightarrow \hat{a}\cdot e^{i\omega_0 t}$ and taking average on initial vacuum state, we can express the classical dynamics of the intra-cavity field as
\begin{equation}
\label{EoMc}
    \langle\dot{\hat{a}}\rangle=-\frac{\gamma}{2}\langle\hat{a}\rangle+E\langle\hat{a}\rangle^*.
\end{equation}
The solution to Eq.~\eqref{EoMc} has the form $\langle\dot{\hat{a}}\rangle_k=S_k\cdot e^{\lambda_kt}$, where $\lambda_k$ and $S_k$ are the $k$th eigenvalue and eigenstate of matrix $E$ \cite{patera2010quantum}. We label the eigenvalue with the largest modulus as $\lambda_0$. As $\lambda_0$ can always be made a real number by adjusting global phase, the criteria for threshold becomes $\lambda_0=\gamma/2$. It is noteworthy that the eigenstates of intra-cavity modes are different from the characteristic modes obtained with Bloch-Messiah decomposition at the output.

\begin{figure}[tb]
\centering
\includegraphics[width=2.5 in]{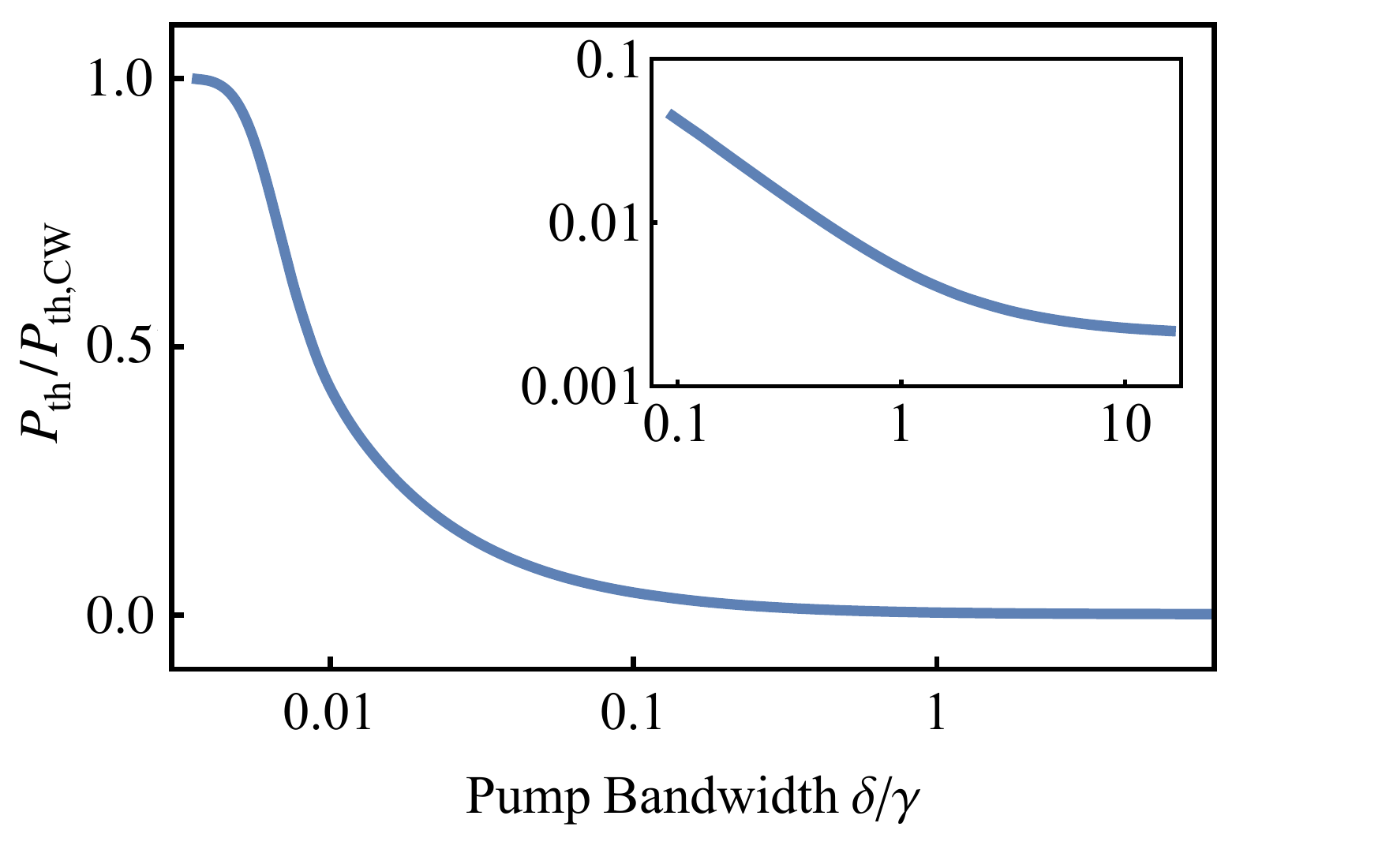}
\caption{\label{fig:PP} The intra-cavity pump threshold power $P_{\rm th}$ with dependence on pump bandwidth $\delta$. The power is normalized with CW intra-cavity threshold $P_{\rm th,CW}=\epsilon\gamma^2/8\kappa^2$, with $\epsilon$ the absolute permittivity. The inset is the log scale plot. This figure assumes $\gamma_p=2\gamma$.}
\end{figure}

\section{Numerical simulation}
As can be seen from the last section, all critical properties of the output state depend on the distribution of squeezing amplitude $\xi_k$. In order to get further insight, the generation of pulsed squeezing in photonic ring cavities is investigated numerically. Without loss of generality, we assume the input pump has Gaussian spectrum shapes $E_p(\omega)\propto e^{-(4\ln{2})(\omega-\omega_0 )^2/\delta^2}$  with full-width-half-maximum (FWHM) $\delta$. The threshold condition needs to be determined first. Using the condition $\lambda_0=\gamma/2$, we obtain the relation between intra-cavity threshold power $P_{\rm th}$ and pump bandwidth $\delta$ (Fig.~\ref{fig:PP}). Here we assume the pump and signal modes have the same quality factor, thus $\gamma_p=2\gamma$. Monotonic decay of the intra-cavity threshold power $P_{\rm th}$ with respect to pump bandwidth $\delta$ can be observed, due to the contribution from multiple pump frequency components. Compared with CW pump, the intra-cavity threshold power can be decreased by three orders of magnitude, making this scheme highly power efficient.

\begin{figure}[tb]
\centering
\includegraphics[width=7.8cm]{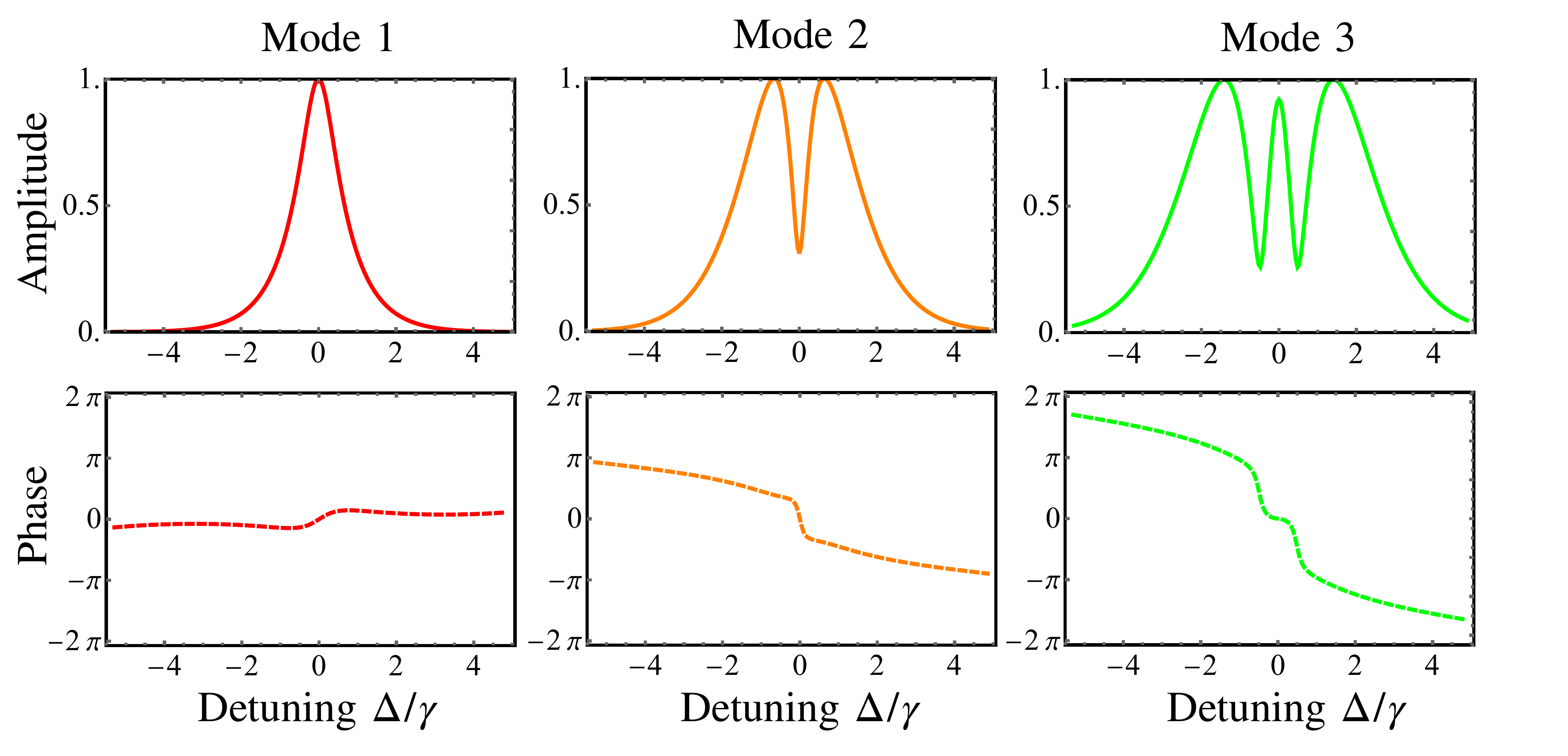}
\caption{\label{fig:MS} The spectral ampliftude (solid) and phase (dash) of the first three characteristic modes near pump threshold. This figure assumes $\gamma_p=2\gamma$ and $\delta=2\gamma_p$.}
\end{figure}

\begin{figure}[tb]
\centering
\includegraphics[width=2.5 in]{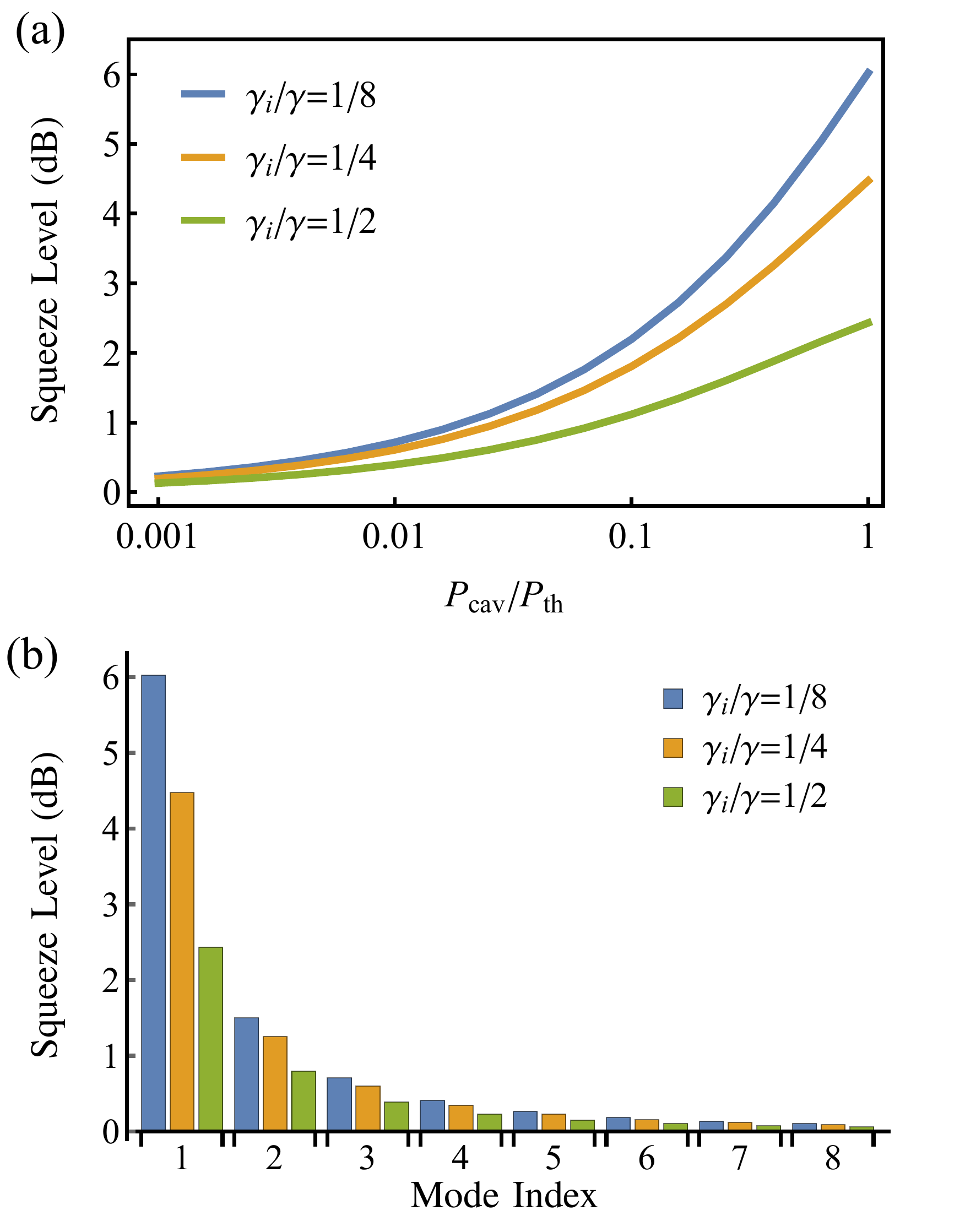}
\caption{\label{fig:SQM} (a) Squeezing level of the first characteristic mode with different intra-cavity power $P_{\rm cav}$ and intrinsic loss $\gamma_i$. (b) Squeezing level of high-order characteristic modes with intra-cavity pump power at $P_{\rm cav}=0.99P_{\rm th}$. This figure assumes $\gamma_p=2\gamma$ and $\delta=2\gamma_p$.}
\end{figure}

\begin{figure}[tb]
\centering
\includegraphics[width=7.8cm]{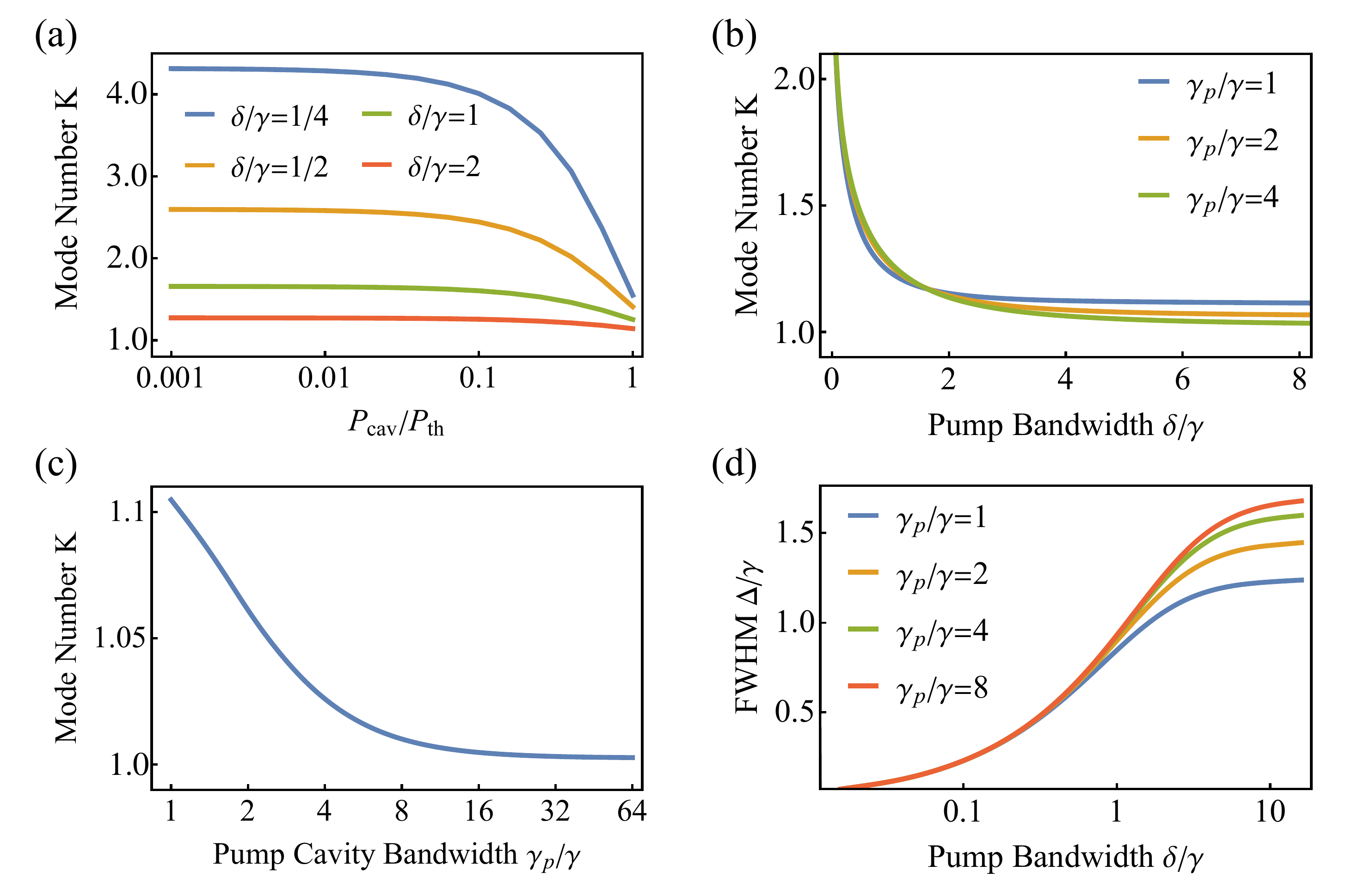}
\caption{\label{fig:K} (a) Effective mode number $K$ as a function of intra-cavity pump power $P_{\rm cav}/P_{\rm th}$ with different pump bandwidth $\delta/\gamma$. This figure assumes $\gamma_p=2\gamma$. (b) Effective mode number $K$ as a function of pump bandwidth $\delta/\gamma$ with different pump cavity bandwidth $\gamma_p$. This figure assumes $P_{\rm cav}=0.99P_{\rm th}$. (c) Effective mode number $K$ as a function of pump cavity bandwidth $\gamma_p$. This figure assumes $P_{\rm cav}=0.99P_{\rm th}$ and $\delta=16\gamma$ (d) FWHM of the first characteristic mode $\Delta$ as a function of pump bandwidth $\delta$ with different pump cavity linewidth $\gamma_p$. This figure assumes $P_{\rm cav}=0.99P_{\rm th}$.}
\end{figure}

Through Bloch-Messiah decomposition of Eq.~\eqref{Decompose}, we can obtain the spectral shape (Fig.~\ref{fig:MS}) and squeezing amplitude (Fig.~\ref{fig:SQM}) of each characteristic mode. Then the squeezing level can be estimated with Eq.~\eqref{Var}. The squeezing level of the first characteristic mode is plotted in Fig.~\ref{fig:SQM}(a). As expected, the squeezing level increases with pump power below threshold, and lower intrinsic loss leads to higher squeezing. We further plot the squeezing level for high-order modes. Due to smaller optical gain,  the squeezing level for high-order modes decrease rapidly (Fig.~\ref{fig:SQM}(b)). Based on Eq.~\eqref{Mode_number}, this indicates the output field will have a small effective mode number and high purity without any filtering and post-selection. When pump power is small, the effective mode number stays constant where entangled photon pairs are generated. With the pump power approaching threshold, the effective mode number drops to a value limited by pump bandwidth (Fig.~\ref{fig:K}(a)).

We further observe that the effective mode number decreases monotonically with both input pump bandwidth $\delta$ and pump cavity linewidth $\gamma_p$ (Fig.~\ref{fig:K}(b)\&(c)). The signal cavity with small linewidth $\gamma$ will function as a spectral-temporal filter to enhance the first characteristic mode and suppress high-order characteristic modes. With larger input pump bandwidth $\delta$ and pump cavity linewidth $\gamma_p$, the filtering effect is more significant, thus leading to smaller effective mode number. This filtering effect is different from adding narrow filters after squeezing generation, as the parametric down-conversion and filtering happen simultaneously in the same cavity. Therefore, no extra loss will be introduced. This filtering effect can be clearly observed in Fig.~\ref{fig:K}(d), where the FWHM of the first characteristic mode increases rapidly with small pump bandwidth, but saturate with large pump bandwidth.

In order to access the maximum squeezing, the spectral-temporal shape of the local oscillator must match the first characteristic mode. Based on the fact that the first characteristic mode shape is critically dependent on the filter effect of the signal cavity (Fig.~\ref{fig:K}(d)), we further design an easy and efficient method for local oscillator shaping. 

\begin{figure}[tb]
\centering
\includegraphics[width=7.8cm]{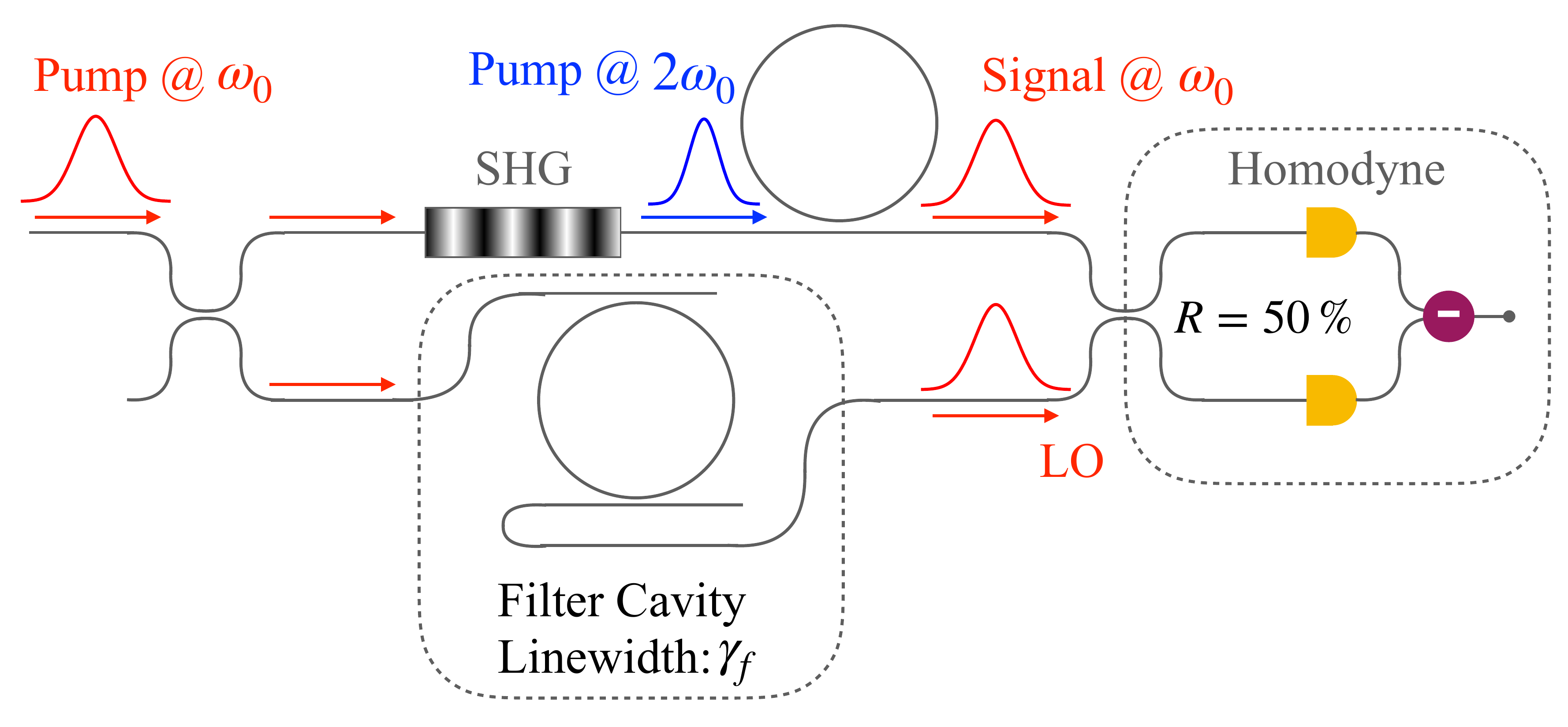}
\caption{\label{fig:ExpDiagram} Proposed setup for local oscillator shaping.}
\end{figure}

\begin{figure}[tb]
\centering
\includegraphics[width=2.5 in]{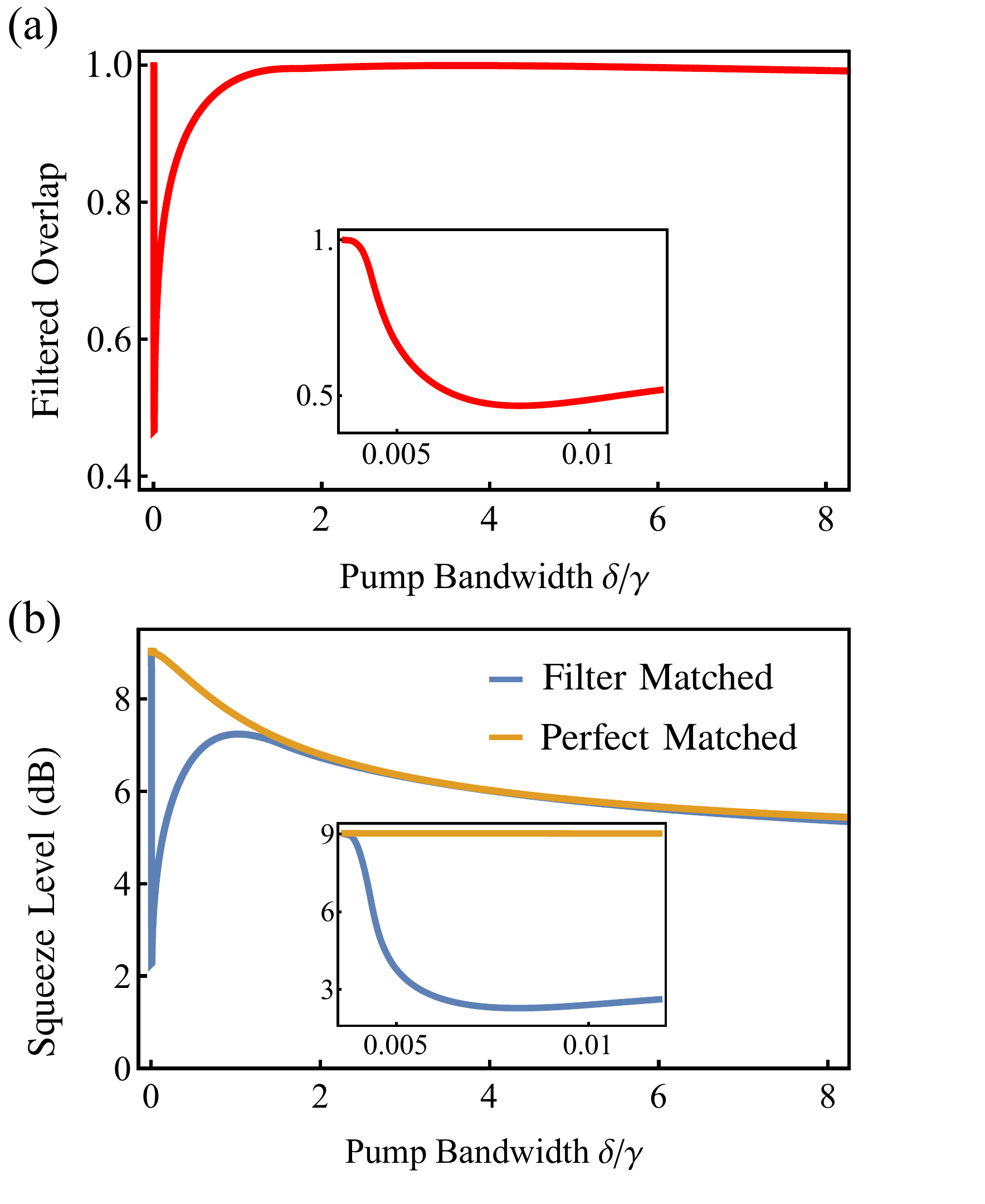}
\caption{\label{fig:OL} (a) The overlap between first characteristic mode and the filtered local oscillator with dependence on $\delta$ (b) Measured squeezing level as a function of $\delta$ with filtered (blue) and perfectly matched local oscillator. Insets are zoom-in of the intermediate region between quasi-CW pump and pulsed pump. This figure assumes $P_{\rm cav}=0.99P_{\rm th}$, $\gamma_p=2\gamma$, and $\gamma_i=1/8\gamma$.}
\end{figure}

As shown in Fig.~\ref{fig:ExpDiagram}, the generation of the pulsed squeezing follows the standard configuration, where a strong optical pulse at the signal frequency is used to generate the pump pulse for parametric down-conversion. A small portion of the optical pulse is tapped to serve as the local oscillator. In order to match the spectral-temporal shape of the first characteristic mode, the optical pulse can simply goes through a optical cavity with Lorentzian lineshape. The cavity linewidth $\gamma_f$ is optimized to obtain the maximum mode overlap and squeezing level (Fig.~\ref{fig:OL}). With small pump bandwidth, the system is in quasi-CW regime, and local oscillator without any mode shaping can achieve near-perfect mode matching. With large pump bandwidth, the filter effect of the signal cavity is significant. A proper filter cavity for the local oscillator is required, and near-perfect matching can be achieved. As small effective mode number is obtained only with large pump bandwidth, this approach for local oscillator shaping should be sufficient.

\section{discussion}
While the current analysis is based on degenerate parametric down-conversion, the generalization to non-degenerate cases is straightforward. The dynamics of intra-cavity modes for signal $\hat{a}_1$ and idler $\hat{a}_2$ can be written as
\begin{equation}
\label{eq:NDME}
\begin{aligned}
    0=&\int d\omega' [i(\omega-\omega_k)-\frac{\gamma}{2}]\delta(\omega-\omega')\hat{a}_k(\omega')\\
    &+\int d\omega' \kappa\varepsilon(\omega+\omega')\hat{a}_l^\dag(\omega')\\
    &+\sqrt{\gamma_i}\hat{a}_k^{(i)}(\omega)+\sqrt{\gamma_c}\hat{a}_{k,in}(\omega)\\
\end{aligned}
\end{equation}
with the index $(k,l)=(1,2)$ or $(2,1)$. Bloch-Messiah decomposition needs to be applied to signal and idler simultaneously. All conclusions for degenerate cases remain valid for non-degenerate cases. The pulsed squeezing generation with non-degenerate configuration can be realized with both parametric down-conversion \cite{guo2017parametric} and spontaneous four-wave mixing \cite{kues2019quantum}. For parametric down-conversion, it is easier to achieve pump cavity linewidth much larger than signal cavity linewidth ($\gamma_p>>\gamma$), due to the vastly different wavelengths. This is beneficial to achieve small effective mode number. Recent development of aluminum nitride \cite{xiong2012aluminum,guo2017parametric,fan2018superconducting,fan2016integrated} and lithium niobate \cite{luo2017chip,wang2018ultrahigh,lu2019periodically} photonics has made it possible to demonstrate pulse squeezing with the proposed method. On the other hand, four-wave-mixing has a wider collection of materials as it does not require non-centro-symmetric crystal structure. Based on Hydex silica glass, non-degenerate four-wave-mixing with pulsed pump is demonstrated for photon pair generation, whose experimental result matches our theoretical calculation \cite{kues2017chip,Exp1}.

\section{Conclusion}
In conclusion, we have proposed a novel approach to generate pulsed squeezing with high temporal purity. Parametric down-conversion in photonic cavities with pulsed pump is analyzed based on Bloch-Messiah decomposition. We show that near-unity effective mode number can be obtained. Large pump cavity linewidth and pump bandwidth are preferred to decrease the effective mode number. As the dependence of effective mode number on pump cavity linewidth and pump bandwidth is monotonic, no delicate balance between the pump power and linewidth is required, making the approach robust. An additional benefit is the low pump threshold due to the contribution from multiple frequency components, leading to the high power efficiency of this approach. We further designed an easy method to realize optimum matching between local oscillator and output characteristic mode for maximum squeezing measurement. The robustness, high efficiency, and easy match of local oscillator make this approach promising for large-scale quantum network and complex quantum state generation. 

\section{Acknowledgments}
This work was supported by Office of Naval Research (N00014-19-1-2190), National Science Foundation (ECCS-1842559, CCF-1907918), and U.S. Department of Energy UT-Battelle/Oak Ridge National Laboratory (4000178321)
\bibliography{Ref}

\end{document}